\documentclass[amsmath,amssymb,superscriptaddress,nobalancelastpage,prb,twocolumn]{revtex4}

\usepackage{graphicx}
\usepackage{color}
\def\Re{\textrm{Re}}
\def\Im{\textrm{Im}}
\def\bfk{\textbf{k}}

\newcommand{\LSCO}{La$_{2-x}$Sr$_{x}$CuO$_4$}
\newcommand{\LSCOov}{La$_{1.83}$Sr$_{0.17}$CuO$_4$ }
\begin{document}

\title{Anisotropic quasiparticle scattering rates in slightly underdoped to
optimally doped high-temperature \LSCO\ superconductors} 
 \author{J.\ Chang}
\affiliation{Laboratory for Neutron Scattering, ETH Zurich
 and Paul Scherrer Institute, CH-5232 Villigen PSI, Switzerland}

 \author{M.\ Shi}
\affiliation{Swiss Light Source, Paul Scherrer Institute,
 CH-5232 Villigen PSI, Switzerland}

\author{S.\ Pailh\'es}
\affiliation{Laboratory for Neutron Scattering, ETH Zurich
 and Paul Scherrer Institute, CH-5232 Villigen PSI, Switzerland}

\author{M.\ M\aa nsson}
\affiliation{Materials Physics, Royal Institute of Technology KTH, S-164 40 Kista, Sweden }

\author{T.\ Claesson}
\affiliation{Materials Physics, Royal Institute of Technology KTH, S-164 40 Kista, Sweden }

\author{O.\ Tjernberg}
\affiliation{Materials Physics, Royal Institute of Technology KTH, S-164 40 Kista, Sweden }

 \author{A. Bendounan}
\affiliation{Laboratory for Neutron Scattering, ETH Zurich
 and Paul Scherrer Institute, CH-5232 Villigen PSI, Switzerland}

 \author{Y. Sassa}
\affiliation{Laboratory for Neutron Scattering, ETH Zurich
 and Paul Scherrer Institute, CH-5232 Villigen PSI, Switzerland}

\author{L.\ Patthey}
\affiliation{Swiss Light Source, Paul Scherrer Institute, CH-5232 Villigen PSI, Switzerland}

\author{N.\ Momono}
\affiliation{Department of Physics, Hokkaido University - Sapporo 060-0810, Japan}

\author{M.\ Oda} 
\affiliation{Department of Physics, Hokkaido University - Sapporo 060-0810, Japan}

\author{M.\ Ido}
\affiliation{Department of Physics, Hokkaido University - Sapporo 060-0810, Japan}

\author{S. Guerrero}
\affiliation{Condensed Matter Theory Group, Paul Scherrer Institute, CH-5232 Villigen PSI, Switzerland}

\author{C. Mudry}
\affiliation{Condensed Matter Theory Group, Paul Scherrer Institute, CH-5232 Villigen PSI, Switzerland}

\author{J.\ Mesot}
\email{joel.mesot@psi.ch}
 \affiliation{Laboratory for Neutron Scattering, ETH Zurich
 and Paul Scherrer Institute, CH-5232 Villigen PSI, Switzerland}
 \affiliation{Institut de la materi\`ere complexe, Ecole Polytechnique Fed\'ed\'erale de Lausanne (EPFL), CH-1015 Lausanne, Switzerland} 
\begin{abstract}
An angle-resolved photoemission study of the scattering rate in the superconducting
 phase of the high-temperature 
superconductor \LSCO\ with $x=0.145$ and $x=0.17$, as a function of binding energy and momentum,
is presented. We observe that the scattering rate scales linearly with
binding energy up to the high-energy scale $E_1\sim0.4$ eV. 
The scattering rate is found to be strongly anisotropic, with a minimum 
along the (0,0)-($\pi,\pi$) direction. 
A possible connection to a quantum-critical point is discussed.

\end{abstract}

\date{\today}
\pacs{74.72.Dn, 79.60.-i}

\maketitle
\section{Introduction}
Angle-resolved photoelectron spectroscopy (ARPES) 
is a powerful probe of electronic 
interactions in solids. 
For example, in studies of high-temperature 
superconductors (HTSC), a low-energy kink,
herein denoted $E_0$, was observed 
 in the (0,0) to ($\pi$,$\pi$) direction (nodal direction)
 of the quasi-particle (QP) spectra.\cite{kaminskiprl00,lanzara01} 
Recently, the high-energy part (0.2-1.5 eV) of the ARPES spectra
 has attracted considerable attention \cite{Grafprl07,ronningprb05,Xie_condmat06,chang_condmat06,pan_condmat06,vallaprl07,meevasana_condmat06}. Graf \emph{et al.} \cite{Grafprl07} reported 
 the existence 
 of two high-energy anomalies $E_1$ and $E_2$ in the nodal dispersion. 
These three anomalies, in the nodal spectra,
 $E_0\approx 0.06$ eV, $E_1\approx0.4$ eV, and 
$E_2\approx0.8$ eV seem to be universal for the cuprates 
and they have been interpreted
in terms of many-body interactions.\cite{Zhu_condmat07,manousakisprb,wangcondmat06,macridincondmat07,markiewiczcondmat07,byczuknatphys}  
We have shown in Ref.~\onlinecite{chang_condmat06} that the high-energy anomaly $E_1$ 
exists throughout
the whole Brillouin zone (BZ) and that 
 $E_1$ disperses continuously as one moves from the 
nodal to the antinodal [(0,0)-$(\pi,0)$] direction.

Transport 
measurements 
 have revealed 
anomalous normal-state (NS) 
properties of optimally doped HTSCs.
The NS 
resistivity, at optimal doping, is found to scale linearly with
the temperature $T$ up to $T \sim$ 1000 K.\cite{takagiprl92} 
This part of the phasediagram is therefore also known as the
 strange metal phase. 
 Although anomalous NS  
properties have  been described 
successfully by the marginal Fermi liquid (MFL)
phenomenology,\cite{varmaprl89} 
there is still
no consensus for the underlying 
interactions responsible for HTSC and
these anomalous 
properties. 

In this letter, we 
investigate
the QP scattering rate 
in \LSCO (LSCO) close to optimally doping. 
 Our main findings, valid for energies much larger 
 than the superconducting gap, are twofold.
 First, the dominant scattering channel scales linearly with the binding energy
$\omega$.
Second, this scattering channel is highly anisotropic,
exhibiting a sharp minimum along the nodal direction.
We emphasize that while these results do not elucidate the pairing mechanism 
of HTSC, they provide constrains to any theory of the strange metal
phase.

 \section{Methods}
 Single crystals 
 LSCO with $x=0.145$ and $x=0.17$
 were grown using the
travelling solvent floating zone method.\cite{Nakano02}
Both samples have a transition temperature $T_c\approx 36$ K with
$\Delta T_c\approx1.5$ K.
The ARPES experiments were performed on the Surface/Interface Spectroscopy (SIS) beamline
at the Swiss Light Source 
of the Paul Scherrer Institute. 
The spectra were acquired with a SCIENTA SES2002 electron analyzer, 
which was  calibrated by recording spectra from polycrystalline 
copper on the sample holder. 
The measurements 
were performed at $T=15$ K under ultra-high vacuum 
using 55
 eV circularly polarized photons with an energy resolution of 
 17-40 meV. Data were recorded in the 2nd BZ, but are 
  presented in the 1st BZ, for convenience. 
 
  \section{Results}
\begin{figure}
\includegraphics[width=0.4\textwidth]{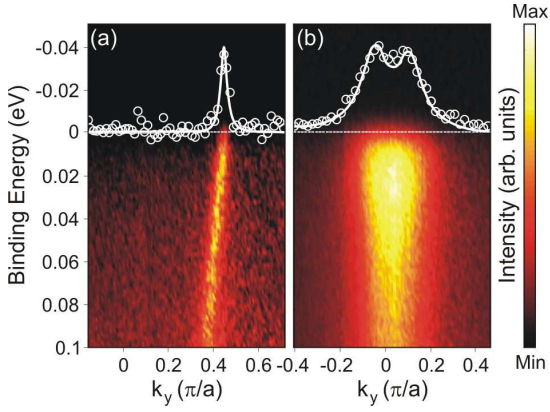}
\caption{(color) (a-b) ARPES intensity, recorded on $x=0.145$, for nodal and anti-nodal cuts, respectively. 
The white points are the MDC at $E_F$.
The intensity ratio between nodal and anti-nodal is $\sim1/3$. 
}\label{fig:fig1a}
\end{figure}

Starting with the low-energy properties, we plot in 
Figs.~\ref{fig:fig1a}(a) and (b) the ARPES intensity as a function of 
binding-energy and electron momentum $\bfk$, along cuts through 
the nodal and anti-nodal  
points, respectively. 
The nodal spectrum is characterized by sharp peaks and the leading edge of
the energy distribution curve (EDC) at $\bfk_F$ reaches the Fermi level $E_F$. 
The MDC linewidth
are much broader in the anti-nodal spectrum and the leading edge of 
the EDCs at $\bfk_F$ is shifted away from $E_F$ due to the presence of an energy
gap $\Delta$.\cite{ming} The double peak structure of anti-nodal MDC linewidth 
stems from the cut crossing two branches of the QP dispersion.

\begin{figure}
\includegraphics[width=0.4\textwidth]{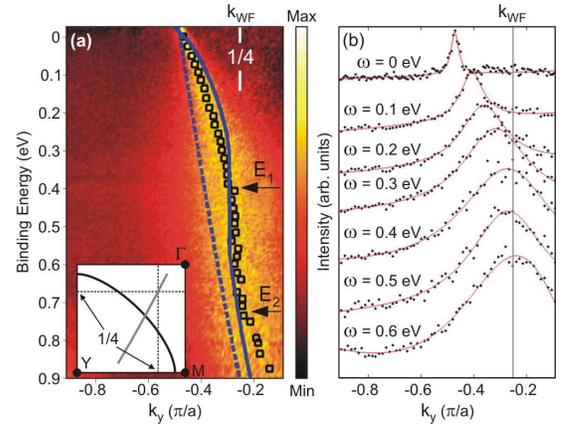}
\caption{(color) (a) MDC-normalized ARPES spectra, recorded
 on $x=0.145$, for the nodal cut shown in the inset.
Black squares represent the MDC peak-positions.  
 Dashed blue line represents
  the bare band dispersion $\varepsilon^{\ }_{\mathbf{k}}$. 
  Solid blue line represents the renormalized dispersion 
  obtained from discussions below.
(b) MDCs for $\omega$
up to $0.6$ eV. The red lines represents  fits to the data with a Lorentzian lineshape on a sloping background.
}\label{fig:fig1}
\end{figure}

\begin{figure}
\includegraphics[width=0.4\textwidth]{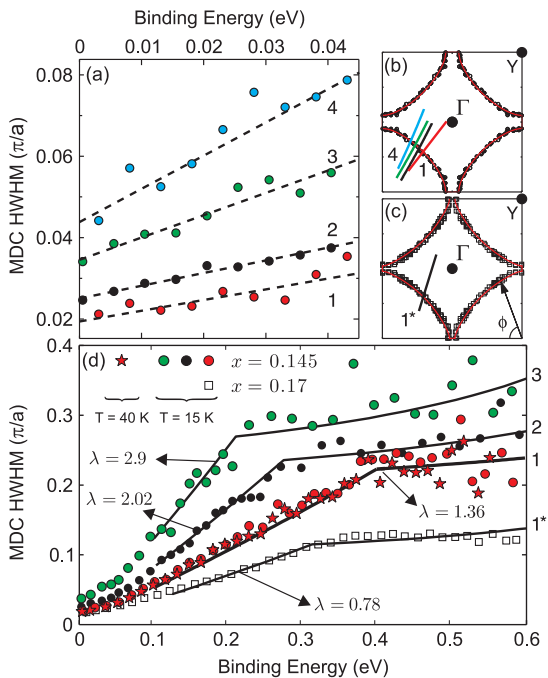}
\caption{(color) (a) Low-energy dependence of the MDC linewidths for the 
cuts shown in (b). Dashed lines represent linear fits 
of the measured scattering rate. (b-c) Fermi surface of $x=0.145$ and
 $x=0.17$ samples, respectively. The red lines are tight-binding fit to the data. (d) $\omega$-dependence of MDC linewidth
 for cuts 1-3 in (b) and the cut in (c).  Cut 1 in (b) was measured
  both in the SC state (circular points) and in the NS (star points) at $T=40$ K.   
 The solid lines are $\Im \Sigma(\omega)/\textit{v}_\phi$ 
 where $\Im \Sigma(\omega)$ and $\textit{v}_{\phi}$ are 
 given in the text. 
 }\label{fig:fig2}
\end{figure}

In Fig.~\ref{fig:fig1}(a) we show the ARPES
 intensity up to very high binding energy 
for the nodal cut shown in the inset. 
The background was subtracted and the intensity
 was normalized to the maximum intensity 
of the momentum distribution curves (MDC)
 for each energy step.
The open black squares
indicate the dispersion extracted from
MDC analysis according to Fig.~\ref{fig:fig1}(b) 
(the blue lines will be explained below).
 As previously reported,\cite{Grafprl07}  
 the nodal spectrum exhibits two high-energy anomalies, $E_1$ and $E_2$,
 as indicated by black arrows in Fig.~\ref{fig:fig1}(a).
 For $E_1<\omega<E_2$
  the MDC peaks are pinned at $\bfk_{WF}\approx(1/4,1/4)$, while
   for $\omega>E_2$ 
   the MDC peaks
 disperse again. 
 These anomalies
 have become known as the waterfall (WF) feature. Herein, the waterfall
 refers only to the $E_1(\phi_{WF})$ anomaly, and we use the 
 notation
 $\bfk_{WF}=(\left|\bfk\right|,\phi)$, with the polar angle
 $\phi$  defined from the Y-point as shown in Fig.~\ref{fig:fig2}(c).
This letter is dedicated to the study of 
the QP scattering rate $\Im\Sigma(\phi,\omega)$
that we model by assuming that it is the product of the MDC linewidth, $\Gamma(\phi,\omega)$,
 and a characteristic velocity $v(\phi)$, to be defined more precisely below.
 The polar angle $\phi$ is here, to a first approximation,
 labeling the cut along which the linewidth is measured.  
 We limit our analysis to $\omega<0.6$ eV  
 where well-defined Lorentzian-shaped peaks, on nearly flat background, are
observed in the MDC, as shown in Fig.~\ref{fig:fig1}(b).
We examine the low- and high-energy dependence of
the half-width at half-maximum (HWHM) $\Gamma(\phi,\omega)$ extracted 
from Lorentzian fits to the MDC from Fig.~\ref{fig:fig2}(a,d).
The Fermi surfaces of  LSCO with $x=0.145$ and $x=0.17$ 
shown in 
Figs.~\ref{fig:fig2}(b) and (c) respectively, are 
consistent
with previous reports.\cite{yoshidaprb06} 
The color code of the cuts in Fig.~\ref{fig:fig2}(b,c) is 
the same as that in Fig.~\ref{fig:fig2}(a,d).

\begin{figure}
\begin{center}
\includegraphics[width=0.42\textwidth]{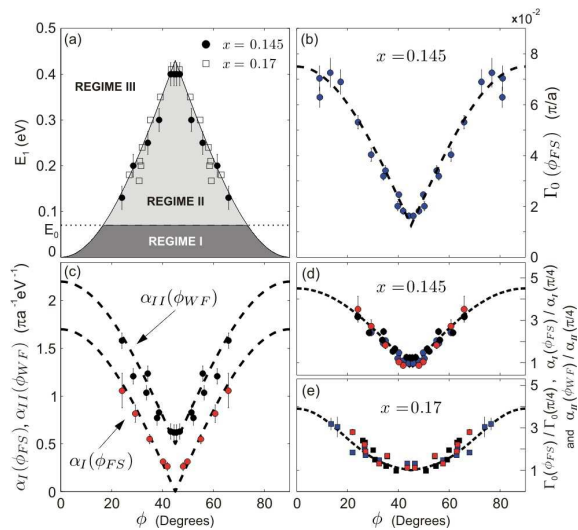}
\caption{(color) (a) Momentum dependence of $E_1$. Solid
 line is obtained from Eq.~(\ref{eq:eq1}). 
(b) MDC linewidth $\Gamma_0$ as a function of the FS angle $\phi_{FS}$.  
(c) Angular dependence of $\alpha_{I}(\phi)$ and
 $\alpha_{II}(\phi)$. (d-e) $\Gamma_0(\phi_{FS})/\Gamma_0(\pi/4)$, 
 $\alpha_{I}(\phi_{FS})/\alpha_{I}(\pi/4)$ and
 $\alpha_{II}(\phi_{WF})/\alpha_{II}(\pi/4)$  for $x=0.145$
  and $x=0.17$, respectively. Blue, red, and black points
 are $\Gamma_0$, $\alpha_I$, and $\alpha_{II}$, respectively. 
  Data have been symmetrized with respect to the
  nodal direction. Dashed lines are guides to the eye. 
}\label{fig:fig3}
\end{center}
\end{figure}

Before studying 
 $\Gamma(\phi,\omega)$,
we first discuss the $\phi$-dependence 
of the high-energy
anomaly $E_1$.  
Figure~\ref{fig:fig3}(a) 
 shows 
$E_1$ extracted from the anomaly 
in the scattering rate shown in Fig.~\ref{fig:fig2}(d). 
$E_1(\phi_{WF})$ 
  disperses strongly and 
   we have previously suggested the 
   following phenomenological form,
    \begin{equation}\label{eq:eq1}
    E_1(\phi_{WF})= E_1(\pi/4)\left[1-|\cos(2\phi_{WF})|\right]
    \end{equation}
with $E_1(\pi/4)=0.43$ eV.\cite{chang_condmat06} 
Within the experimental uncertainty, there is no 
significant difference between $E_1(\phi_{WF})$
for LSCO with $x=0.145$ and $x=0.17$. 

The energy scales $E_0$ and $E_1$
define three distinct characteristic 
regimes shown in Fig.~\ref{fig:fig3}(a).  
Regime I is the low-energy regime 
$E_F<\omega<\min\{E_0,E_1(\phi_{WF})\}$, 
followed by an intermediate 
regime II defined by  $E_0<\omega<E_1(\phi_{WF})$. 
Finally, we define the high-energy regime 
III by $E_1(\phi_{WF})<\omega$.

Although the main purpose of this paper is 
to study the QP scattering rate $\Sigma(\phi,\omega)$ in regime II 
we present $\Gamma(\phi,\omega)$ in  the three  regimes.
The MDC linewidth $\Gamma(\phi,\omega)$  in regimes I and II  obey
 \begin{equation}\label{eq:eq2}
 \Gamma_{i}(\phi,\omega)=\Gamma^0_i(\phi)+\alpha_{i}(\phi)\omega, \qquad i=\hbox{I,II}.
 \end{equation}
 Consistent with previous ARPES \cite{yoshidajphys}
  and transport \cite{narduzzo} measurements
the elastic term $\Gamma^0_I(\phi_{FS})$ is 
 anisotropic as shown in Fig.~\ref{fig:fig3}(b).
 The parameter $\alpha_I(\phi_{FS})$, related to the inelastic scattering,
    is analyzed by linear fits to the scattering rate 
  $\Gamma_I(\phi_{FS},\omega)$, see dashed lines in Fig.~\ref{fig:fig2}(a).  
  We show in Fig.~\ref{fig:fig3}(c) the 
  $\phi_{FS}$-dependence of $\alpha_I(\phi_{FS})$ in the vicinity of the nodal point. 
  The linear dependence of $\Gamma_I(\phi_{FS},\omega)$ 
  was also observed in Bi2212 \cite{kaminskiprb05,vallaprb06,yamasakiprb07} and interpreted  in ref. 
  \cite{yamasakiprb07} as a signature of the d-wave nodes.

 We now turn to regime II for which $\Gamma_{II}^0(\phi)$ is 
 negligible for $x=0.145$ and 
 the angular dependence of $\alpha_{II}(\phi_{WF})$
 is shown in Fig.~\ref{fig:fig2}(c). Observe  that the coefficient $\alpha_{II}$ is 
 the same for both  $T_c>T=15$ K and  $T_c<T=40$ K, see Fig.~\ref{fig:fig2}(d).
 This is expected since the relevant energy scale 
 in regime II is an order of magnitude larger than 
 the maximum of the superconducting gap.
 Hence the linear dependence on $\omega$ in Eq.~(\ref{eq:eq2})
 cannot be attributed to the d-wave nodes. 
 Nevertheless and remarkably 
 $\Gamma^0_I$, $\alpha_I$, and $\alpha_{II}$,  
 follow a very similar angular dependence. 
To show this, we plot $\Gamma_0(\phi_{FS})/\Gamma_0(\pi/4)$, $\alpha_{I}(\phi_{FS})/\alpha_{I}(\pi/4)$ and $\alpha_{II}(\phi_{WF})/\alpha_{II}(\pi/4)$ in Fig.~\ref{fig:fig3}(d). For 
\LSCOov\ we find the same anisotropic dependence although with a slightly weaker and
flatter dependence around the nodal direction, see Fig.~\ref{fig:fig3}(e). 

The approximate Lorentzian shape of the MDCs suggests that 
one can neglect the $\bfk$- and $\omega$-dependence of the
photoelectron matrix elements. 
If so, we can approximate the ARPES intensity by 
the single-particle spectral function 
  \begin{equation}\label{eq:eq4}
  A(\bfk,\omega)=\frac{1}{\pi}\frac{-\Im\Sigma(\bfk,\omega)}{\left[\omega-\Re\Sigma(\bfk,\omega)-\varepsilon_\bfk\right]^2+[\Im\Sigma(\bfk,\omega)]^2}.
  \end{equation}
Here $\Sigma(\bfk,\omega)$ is the self-energy 
 and $\varepsilon_\bfk$ is the bare band dispersion. 
We model   $\varepsilon_\bfk$ with the tight-binding dispersion
\begin{equation}\label{eq:eq5}
\begin{split}
\varepsilon^{\ }_{\mathbf{k}}=&-2t[\cos k_x a+\cos k_y a]-4t'\cos k_x a \cos k_y a \\
&-2t''[\cos 2 k_x a +\cos 2k_y a]-\mu,
\end{split}
\end{equation}
 where
  $\mu$ is the chemical potential, and $t$, $t^{\prime}$, 
  and $t^{\prime\prime}$ denote nearest, second-nearest, 
  and third-nearest neighbor hopping
  integrals on a square lattice, respectively. The ratios $\mu/t$, $t^{\prime}/t$,
   and $t^{\prime\prime}/t$, given in table \ref{tab:tab1},
    are chosen such that $\varepsilon^{\ }_{\mathbf{k}}=0$
     fits the experimentally determined Fermi surfaces,
      see Figs.~\ref{fig:fig2}(b) and (c). 
      Assuming that the bandwidth $t$ varies slowly 
      within the doping range of interest, we use
      for the 
  bare band $\varepsilon^{\ }_{\mathbf{k}}$
   (see dashed blue line in Fig.~\ref{fig:fig1}) 
   $t=0.48$ eV  for both $x=0.145$ and $x=0.17$.\cite{pavariniprl01}

    In regime II we  analyze the cuts shown in Fig.~\ref{fig:fig2}(b-c) 
  with a generalized marginal Fermi liquid (MFL) self-energy 
    \begin{equation}\label{eq:eq6}
 \Im \Sigma(\phi_{WF},\omega)= \frac{-\lambda(\phi_{WF}) \pi}{2} \begin{cases} |\omega|,& |\omega|<\omega_c(\phi_{WF}), \\ \omega_c(\phi_{WF}),& |\omega|>\omega_c(\phi_{WF}), \end{cases}
\end{equation}
and
\begin{equation}\label{eq:eq7}
\Re \Sigma(\phi_{WF},\omega)=- \lambda(\phi_{WF}) \left[\omega\ln\left(\frac{\omega_c(\phi_{WF})}{\omega}\right)+...\right].
\end{equation}
The conventional MFL ansatz \cite{Zhu_condmat07,varmaprl89}  for the self-energy assumes that 
the dimensionless coupling $\lambda$ and the characteristic 
energy cutoff $\omega_c$ are $\phi$-independent.
Motivated by Eqs.~(\ref{eq:eq2}), 
we are going to relax this assumption
in order to describe the MDC linewidth of Fig.~\ref{fig:fig2} from Eqs.~(\ref{eq:eq4}-\ref{eq:eq7}). 
Along the cuts shown in Fig.~\ref{fig:fig2}(b-c), the MDCs
have a Lorentzian shape with HWHM
 $\Gamma(\phi,\omega)=\Im \Sigma(\phi,\omega)/\textit{v}_{\phi}$ 
 where $\textit{v}_\phi=d\varepsilon_\bfk/d\bfk$  is the velocity along the cut.\cite{vallascience99}
 Combining Eqs. (\ref{eq:eq2}) and (\ref{eq:eq6}), it then follows that
 \begin{equation}\label{eq:eq8}
 \pi\lambda(\phi_{WF})=\alpha_{II}(\phi_{WF})\textit{v}_{\phi}\approx \alpha_{II}(\phi_{WF})\textit{v}_{\phi_{WF}}.
 \end{equation}
 This approximation is valid in the vicinity 
  of the nodal point where
the bare-band velocity $\textit{v}_{\phi}$ is 
weakly dependent on $\bfk$ for $\omega<0.6$ eV but breaks down upon approaching 
the van Hove singularity of $\varepsilon^{\ }_{\mathbf{k}}$ in the anti-nodal region.
Second, we approximate the cutoff energy by 
\begin{equation}
\omega_c(\phi_{WF})\approx E_1(\phi_{WF}).
\end{equation}
Now, the renormalized dispersion is the solution of
$\omega_p(\bfk)=\Re\Sigma(\omega_p(\bfk))+\varepsilon_\bfk$.
In this fashion we obtain a consistent agreement for both the 
renormalized dispersion [solid blue line  in Fig.~\ref{fig:fig1}(a)] and 
the MDC linewidth [solid lines in Fig.~\ref{fig:fig2}(d)]. 
  Thus, in contrast 
to earlier claims,\cite{inosov_condmat07} we have shown that the 
WF features can 
be described by a Kramers-Kronig consistent 
self-energy
function $\Sigma(\phi,\omega)$. 
We would like to stress that the
 $\alpha_{II}(\phi_{WF})$ and  $\textit{v}_{\phi_{WF}}$
dependencies on $\phi_{WF}$ do not cancel out, leaving a net anisotropic
coupling parameter $\lambda(\phi_{WF})$.
Furthermore, the observation that $\alpha_{II}(\phi)$ has a stronger dependence on
doping than $\varepsilon_{\bfk}$ implies that the coupling constant 
$\lambda(\phi_{WF})$ decreases with 
overdoping.

\section{Discussion}
We have shown that both the elastic and inelastic scattering rates are highly
anisotropic. However, an isotropic channel may be hidden 
by the dominant anisotropic scattering channel. 
Recently, two scattering channels have been identified, in the overdoped
regime of \textit{ Tl$_2$Ba$_2$CuO$_{\delta+6}$ (Tl2201)}, 
by an angular magnetoresistance oscillation (AMRO) study.\cite{abdeljawad}
One channel, related to electron-electron scattering, 
is isotropic and exhibits  
$T^2$-dependence.
A second channel, of unknown origin, is anisotropic and depends linearly on $T$.
For even more overdoped samples, resistivity measurements on \LSCO\
have demonstrated that the electron-electron scattering channel
is completely dominant.\cite{nakamaeprb,abdeljawad}
\begin{table}
\caption{\label{tab:tab1} Tight binding parameters for \LSCO.}
\begin{ruledtabular}
\begin{tabular}{cccccc}
    Compound & $t$ [eV]& $\mu/t$&$t^\prime/t$& $t^{\prime\prime}/t$& $E_1(\pi/4)/t$ \\\hline
    $x=0.145$ & 0.48& 0.68 & -0.125& 0.078& 0.9 \\
   $x=0.17$ & 0.48& 0.84& -0.144 & 0.072& 0.9\\
\end{tabular}
\end{ruledtabular}
\end{table}

   The picture that emerges from this work and previous transport
 measurements \cite{abdeljawad,nakamaeprb} is the 
following. In the underdoped regime the dominant scattering channel is 
highly anisotropic and exhibits MFL behavior. 
 Upon further hole doping
this channel gradually decreases and conventional electron-electron interactions
become increasingly important. Eventually in the extremely overdoped regime, 
$x>0.3$, electron-electron interactions are the dominant scattering mechanism. 

\section{Conclusion}
In conclusion, we have presented
a comprehensive study of the $\omega$- and $\bfk$-dependence of 
the scattering rate in the vicinity of optimally doped \LSCO.
The dominant inelastic scattering channel scales linearly with
$\omega$ up to the onset of the waterfall feature.
Remarkably, both the elastic and inelastic scattering channels are 
strongly anisotropic, with minima along the nodal direction. 
This anisotropic MFL behavior can be used to discriminate between
competing theories for the strange metallic phase in high-temperature 
superconductors that rely on the single-band Hubbard model,\cite{kakehashi05}
the existence of a quantum critical point,\cite{varmaprl99,chakravartyprb01,kimnematic,sachdev} 
or the separation of spin and charge quantum numbers.\cite{PALee06,andersonnatphys06}

\section{Acknowledgment}
This work was supported by the Swiss NSF 
(through NCCR, MaNEP, and grant Nr 200020-105151), the Ministry
 of Education and Science of Japan and the Swedish Research Council.
  This work was performed at SLS of the Paul Scherrer Institute, 
  Villigen PSI, Switzerland. We thank the beamline staff of X09LA
   for their excellent support and
  we are grateful to Lijun Zhu and Chandra Varma for enlightening discussions.


\end{document}